\documentclass[12pt]{article}
\usepackage{graphicx}
\usepackage{amsfonts}
\usepackage{epsfig}
\usepackage{epstopdf}
\usepackage{jheppub}
\usepackage{amsmath}
\newcommand{\be}{\begin{equation}}
\newcommand{\ee}{\end{equation}}
\def\bea{\begin{eqnarray}}
\def\eea{\end{eqnarray}}

\begin{document}

\title{Universal Bounds in Even-Spin CFTs}
\author{Joshua D. Qualls}
\affiliation{Department of Physics,
National Taiwan University, Taipei, Taiwan}
\emailAdd{joshqualls@ntu.edu.tw}

\abstract{ We prove using invariance under the modular $S-$ and $ST-$transformations that every unitary two-dimensional conformal field theory (CFT) having only even-spin primary operators (with no extended chiral algebra and with right- and left-central charges $c,\tilde{c}>1$) contains a primary operator with dimension $\Delta_1$ satisfying $0<\Delta_1<\frac{c+\tilde{c}}{24}+0.09280...$. After deriving both analytical and numerical bounds, we discuss how to extend our methods to bound higher conformal dimensions before deriving lower and upper bounds on the number of primary operators in a given energy range. Using the AdS$_3$/CFT$_2$ dictionary, the bound on $\Delta_1$ proves the lightest massive excitation in appropriate theories of 3D matter and gravity with cosmological constant $\Lambda<0$ can be no heavier than 1/$8G_N+O(\sqrt{-\Lambda})$; the bounds on the number of operators are related via AdS/CFT to the entropy of states in the dual gravitational theory. In the flat-space approximation, the limiting mass is exactly that of the lightest BTZ black hole. 
}

\maketitle

\section{Introduction}

The problem of quantizing general relativity is one of the outstanding questions of theoretical physics. In order to better understand the gravity of our four-dimensional universe, we often consider the lower-dimension model of quantum gravity in 2+1 dimensions. Decades of effort have given important insight into quantum gravity in 2+1 dimensions \cite{1p,2p,3p,4p,5p}. Studies of maximally symmetric anti-de Sitter (AdS) spaces for the case of negative cosmological constant have been particularly successful.  It has been known since the work of Brown and Henneaux \cite{26p} that quantum gravity in (2+1)-dimensional spacetime is related to two-dimensional conformal field theory. This connection is a specific case of the AdS/CFT correspondence. This correspondence relates quantum gravity in an asymptotically AdS spacetime to a conformal field theory in one lower dimension, which can regarded as living at the spatial boundary of the AdS spacetime.  Using this correspondence allows for problems in quantum gravity to be addressed using the tools of conformal field theories.  

Though this duality has numerous and diverse applications, we are interested primarily in the correspondence between the mass of a state in the bulk theory and the conformal dimension of the associated operator in the boundary CFT. By deriving bounds on conformal dimensions in the CFT, we are able to make statements about the allowed mass of states in any corresponding theory of quantum gravity. Of course, two-dimensional conformal theories are interesting in their own right; they describe string worldsheets, phase changes, and many other interesting physical phenomena. It is for all of these reasons and others that we focus our efforts on proving universal bounds on conformal dimensions in two-dimensional conformal field theories.
 
Much work has already been done in deriving bounds on conformal dimensions. The paper \cite{6p} (based on \cite{9p,10p}) examines the gravitational duals of 2D CFTs for which the partition function is holomorphically factorized as a function of the complex structure $\tau$ of the torus. In this class of CFT it can be shown that the lowest primary operator is either purely left- or right-moving, and can have a dimension no larger than $1 + \mbox{min}(\frac{c}{24},\frac{\tilde{c}}{24})$, where $c,\tilde{c}$ are the left-,right-central charge. For all positive integer values of $(\frac{c}{24},\frac{\tilde{c}}{24})$, there exists a unique partition function for which this bound is saturated (though it is unclear if this partition function correspond to an actual conformal field theory \cite{33p}). Other work \cite{7p} considers the case of theories with extended (2,2) supersymmetry. This property allows the authors to exploit the holomorphic dependence on the complex structure without assuming holomorphic factorization. Study of a certain subclass of (2,2) SUSY CFTs suggests a bound that goes as $\Delta_1\leq\frac{c}{24}$ for large central charge. 
 
A more recent paper \cite{hell} derives a general upper bound on the conformal dimension of the lowest primary operator in a general two dimensional conformal field theory, assuming only unitarity, a discrete operator spectrum, and invariance of the partition function under the modular $S$-transformation\footnote{The conclusions also apply to CFTs with continuous spectra that can be realized as limits of CFTs with discrete spectra. For example, the moduli space of the D1-D5 CFT \cite{34p,35p, 36p,37p,38p}.}. The proof does not assume any special properties like holomorphic factorization or supersymmetry, nor does it refer to the bulk three-dimensional spacetime or asymptotic expansion at large central charge. The resulting upper bound is
\begin{equation}
\Delta_1 \leq \frac{c_{\rm tot}}{12} + 0.4736... \label{eq:hell}
\end{equation}
Using the AdS/CFT dictionary, equation (\ref{eq:hell}) translates directly into an upper bound on the mass of the lightest massive state in a theory of gravity in three dimensions.  

Building on this work, the paper \cite{fried} investigated additional constraints from $S$-invariance systematically. The authors applied the next several higher order differential constraints using the linear functional method and found that for finite $c_{\rm tot}$ the bound can be lowered somewhat. In \cite{mine}, the methods of \cite{hell} are generalized to find upper bounds for general conformal dimension $\Delta_n$. This work also provides a lower bound on the number $N$ of primary operators satisfying eq. (\ref{eq:hell}) going as
\begin{equation}
\log N \gtrsim \frac{\pi c_{\rm tot}}{12}. \label{eq:lowernumber}
\end{equation}
An alternate proof of this fact was found in \cite{hart}. The authors considered two-dimensional CFTs with large central charge and a sparse light spectrum and showed $S$-invariance implies both that the free energy is universal for all temperatures and that the microscopic spectrum matches the Cardy entropy for all $\Delta \geq c_{\rm tot}/12$.
 
In this paper, we consider 2D CFTs that are invariant under the $S-$ and $ST-$ transformations. We derive a smaller upper bound on the weight of the lowest primary operator by considering a general two dimensional conformal field theory consisting of only even-spin operators. As in \cite{hell}, we will assume that the Hilbert space has a positive definite norm (ncessary for a consistent interpretation of quantum mechanics), and that the spectrum of operator dimensions is discrete (necessary for well-behaved thermodynamic properties).
By restricting to theories having primary operators with only even spins ($J_A=0, \pm 2...$), we find that $\Delta_1\leq\frac{c_{\rm tot}}{24}+O(1)$. We briefly address how to extend this proof to larger conformal dimensions (as in 
\cite{mine}), before discussing upper and lower bounds on the number of primary operators in a given energy range (as in \cite{hart}). We conclude by investigating the gravitational interpretation of our results. The upper bound on $\Delta_1$ translates directly into an upper bound on the mass of the lightest massive state in a theory of gravity and matter in three dimensions subject to the same even-spin condition. We will see that in the flat space approximation, this limiting mass is precisely that of the lightest BTZ black hole. 

\section{Constraints from $ST$-invariance}

In this section, we review how unitarity and modular invariance lead to constraints on the conformal dimensions of a 2D CFT. The techniques described in this section were developed in \cite{11p,12p}, where they were used to estimate dimensions of operators in special cases, as well as \cite{hell,fried,mine} in deriving upper bounds. Related techniques have been used to bound certain operator dimensions in conformal field theories in various other dimensions \cite{15p, 16p,17p,18p,19p,20p,21p,22p,225p,super1}. 

\subsection{Modular invariance}

We consider a general CFT in two dimensions with positive norm and discrete spectrum. When the spatial direction $\sigma^{1}$ is compactified on a circle of length $2\pi$, the partition function of the theory at temperature $\beta^{-1}$ is given by
\begin{equation}Z(\beta)=\mbox{Tr}\left(\,\exp\{-\beta H\}\,\right),\end{equation}
where $H$ is the Hamiltonian on a circle of length $2\pi$. The partition function can be refined by adding a potential $K^{1}$ for momentum $P_{1}$ in the compact spatial direction:
\begin{equation}Z(\beta,K^{1})\equiv \mbox{Tr}(\exp\{iK^{1}-\beta H\})\end{equation}
Defining the modular parameter $\tau \equiv (K^{1}+i\beta)/2\pi$, as well as the usual modular variable $q\equiv e^{2\pi i \tau}$, the partition function can then be expressed in the form
\begin{equation}Z(\tau,\bar{\tau})\equiv \mbox{Tr}(q^{L_{0}-\frac{c}{24}}\bar{q}^{\tilde{L}_{0}-\frac{\tilde{c}}{24}}).
\end{equation}
Here $c$ and $\tilde{c}$ are the right- and left-moving central charges, and $L_{0}=\frac{1}{2}(H+P_{1})+\frac{c}{24}$, $\tilde{L}_{0}=\frac{1}{2}(H-P_{1})+\frac{\tilde{c}}{24}$ are the right- and left-moving conformal weight operators which fit into the usual Virasoro algebra. The Virasoro generators obey the usual Hermiticity condition $L^{\dagger}_{m}=L_{-m}$, and it follows from unitarity that every primary operator has nonnegative weight, with weight zero if and only if the operator is the identity.

The partition function can be realized as the path integral of the conformal field theory on a torus of complex structure $\tau$ without operator insertions. Large coordinate transformations of the torus have the structure of the modular group PSL($2,\mathbb{Z}$), with the generator $\left( \begin{array}{cc} a & b \\ c & d \\ \end{array}\right)$ acting as $\tau\rightarrow\frac{a\tau+b}{c\tau+d}$. The group is generated by the transformations $T=\left( \begin{array}{cc} 1 & 1 \\ 0 & 1 \\ \end{array}\right)$ and $S=\left( \begin{array}{cc} 0 & -1 \\ 1 & 0 \\ \end{array}\right)$, which act as $\tau\rightarrow\tau+1$ and $\tau\rightarrow -\frac{1}{\tau}$, respectively (such that $S^{2}=-1$ and $(ST)^{3}=1$).
Invariance of the partition function under the T transformation is equivalent to the condition that every state have $h-\tilde{h}\in Z$, where $h,\tilde{h}$ are the state's eigenvalues under $L_{0},\tilde{L}_{0}$. Consequences of invariance of the partition function $Z(\tau,\bar{\tau})$ under the modular $S$-transformation have been studied in depth (e.g.,\cite{19p,mine,fried, hart}).  We turn our attention to consequences of invariance of the partition function under the $ST$-transformation.

\subsection{Intermediate Temperature Expansion}

The following discussion closely follows the derivation given in \cite{hell}, though it has been adapted here to the case of invariance under the $ST$-transformation. We refer the reader there for additional details.

In order to study $ST$-invariance, we focus on the complex modular parameter at the value $\tau=\omega\equiv-1/2 + i\sqrt{3}/2$ such that the point $\omega$ is fixed under the modular transformation $ST: \tau \rightarrow -\frac{1}{\tau+1}$. We have chosen this value of the complex structure in order to be definite---considering the complex conjugate $\bar{\omega}$ (invariant under the modular transformation $(ST)^2$) gives no additional information. We choose a neighborhood of $\tau=\omega$ to parametrize this neighborhood conveniently
\begin{equation}
\tau=\omega e^{s}\approx\omega(1+s). 
\end{equation}
This parameterization is not optimal, as it will not manifestly exhibit $ST$-invariance to all orders. A good parameterization would involve the modular $j$-invariant
\begin{equation}
j(\tau) = 32 \frac{(\theta_2(0;q)^8 + \theta_3(0;q)^8 + \theta_4(0;q)^8)^3}{(\theta_2(0;q) \theta_3(0;q) \theta_4(0;q))^8} ,
\end{equation}
where the $\theta$ are auxiliary theta functions.
This complicated analysis is unnecessary, however---we will only require a constraint at linear order, and the simpler exponential parameterization is therefore sufficient.

Under the $ST$-transformation, $s \rightarrow \omega^{2}s$ near the fixed point $\omega$. Invariance of the partition function under this transformation then tells us that 
\begin{equation}Z(\omega e^{s},\bar{\omega}e^{\bar{s}})=Z(\omega e^{\omega^{2}s},\bar{\omega}e^{\bar{\omega}^{2}\bar{s}})\end{equation}
Scaling $s\rightarrow 0$ and examining the behavior of the partition function is what we shall refer to as the \emph{intermediate temperature expansion}. This terminology is inspired by the ``medium temperature expansion'' discussed in \cite{hell}. Taking successive derivatives evaluated at $s=0$, we see that
\begin{equation}
\left(\frac{\partial}{\partial s}\right)^{N_{L}}  \left(\frac{\partial}{\partial \bar{s}}\right)^{N_{R}} Z\bigg|_{s,\bar{s}=0}=0,\;\;N_L\mbox{ mod 3}\neq N_R\mbox{ mod 3}
\end{equation}
In terms of the parameter $\tau$, this is
\begin{equation}
\left(\tau\frac{\partial}{\partial \tau}\right)^{N_{L}}  \left(\bar{\tau}\frac{\partial}{\partial \bar{\tau}}\right)^{N_{R}} Z(\tau,\tilde{\tau})\bigg|_{\tau=\omega}=0,\;\;N_L\mbox{ mod 3}\neq N_R\mbox{ mod 3} \label{eq:tauconstraint}
\end{equation}
The condition on $N_L$ and $N_R$ reflect the fact that the $ST$-transformation satisfies $(ST)^3:\tau\rightarrow\tau$.

As will be demonstrated below, terms in the partition function have dependence going as $Z\sim e^{-\beta\Delta}e^{iK^1 J}$, where $\beta=-i\pi(\tau-\bar{\tau})$, $K^1=\pi(\tau+\bar{\tau})$, $\Delta$ is the conformal dimension of a state, and $J$ is its conformal spin. The differential constraints given in terms of $\tau$ and $\bar{\tau}$ acting on terms of this form generically lead to complex polynomials and alternating sums. These alternating sums do \emph{not} lead to positivity conditions and are useless for our methods of proof. In order to end up with useful polynomial constraints, we thus need to express the above differential constraints in terms of $\beta$. In general, the constraints (\ref{eq:tauconstraint}) can not be written solely in terms of $\beta$ derivatives; we are able, however, to obtain the lowest order differential constraint
\begin{equation}
\left(\beta\frac{\partial}{\partial\beta}\right)Z(K^1,\beta)\bigg|_{\tau=\omega}=0.
\end{equation}
This constraint from $ST$-invariance corresponds to the lowest-order constraint from $S$-invariance given in \cite{hell},
\begin{equation}
\left(\beta\frac{\partial}{\partial\beta}\right)Z(K^1,\beta)\bigg|_{\tau=i}=0.
\end{equation}
We will use both of these results in the work that follows.

\subsection{Polynomial Constraint}

We will consider the same class of theories as \cite{hell}. In particular, we will consider only theories with $c,\tilde{c}>1$; compact, unitary CFTs with $c\leq 1$ are completely classified and we can inspect the operator spectra directly (see \cite{43p}). We assume the theory has no chiral algebra beyond the Virasoro algebra in order to simplify our analysis (this assumption can be removed to obtain more general results at the expense of weaker bounds---the extension is straightfoward, but nontrivial \cite{mine3}). Using cluster decomposition, we can therefore split our partition function $Z(\tau)$ into a sum over conformal families:
\begin{equation}
Z(\tau)=Z_{id}(\tau)+\sum_{A}Z_{A}(\tau)
\end{equation}
where $A$ refers to the $A^{th}$ primary having conformal weights $h_A$ and $\tilde{h}_A$.
Using a well-known Virasoro representation structure theorem \cite{14p,23p,24p}, we can express the partition function as
\begin{equation}
Z(\tau) = q^{(-c/24)}\tilde{q}^{-\tilde{c}/24} \left[ \prod_{m=1}^{\infty}(1-q^{m})^{-1}\right] \left[\prod_{n=1}^{\infty}(1-\tilde{q})^{-1}\right]\left[(1-q)(1-\tilde{q})+Y(\tau)\right]
\end{equation}
where
\begin{equation}
Y(\tau)=\sum_{A=1}^{\infty}q^{-h_{A}}\tilde{q}^{-\tilde{h}_{A}}
\end{equation}

By introducing conformal dimension $\Delta_{A}\equiv h_{A}+\tilde{h}_{A}$ and conformal spin $J_{A}\equiv h_{A}-\tilde{h}_{A}$, we can express the partition function over primaries as 
\begin{equation}
Y(\tau,\bar{\tau})=\sum_{A=1}^{\infty}e^{-\beta \Delta_{A}}e^{i K^{1}J_{A}}= \sum_{A=1}^{\infty}e^{i \pi (\tau-\bar{\tau}) \Delta_{A} }e^{i \pi (\tau+\bar{\tau}) J_{A}}.
\end{equation}
At this point it is apparent that terms in the partition function have the dependence claimed earlier.
Finally, we can simplify the prefactor. Defining $\hat{E}_{0}\equiv E_0+\frac{1}{12} = -\frac{c+\tilde{c}}{24} + \frac{1}{12} \equiv \frac{2-c_{\rm tot}}{24}$ and $\Delta c\equiv -\frac{c-\tilde{c}}{24}$, we find
\begin{equation}
 q^{\frac{-(c-1)}{24}}\tilde{q}^{\frac{-(\tilde{c}-1)}{24}} = e^{-\beta\hat{E}_0}e^{iK^{1}\Delta c}=e^{i \pi (\tau-\bar{\tau})(E_{0}+\frac{1}{12})} e^{i \pi (\tau+\bar{\tau}) \Delta c}
\end{equation} 
This gives for the full partition function
\begin{gather}
Z(K^1,\beta) = e^{-\beta\hat{E}_0}e^{iK^{1}\Delta c} |\eta(\tau)|^{-2}[(1-q)(1-\tilde{q})+Y(\tau,\bar{\tau})]\nonumber \\
= M(\tau,\bar{\tau})Y(\tau,\bar{\tau})+B(\tau,\bar{\tau}),
\end{gather}
where $M$ and $B$ are defined for convenience.

In what follows it will also be convenient to define some polynomials. We define $g(z)$ by the equation
\begin{equation}
(\beta \partial_\beta)M(\beta)Y(\beta) \bigg|_{\beta=\pi\sqrt{3}} = -|\eta(\omega)|^{-2} \sum_{A=1}^{\infty} e^{-\pi\sqrt{3} (\Delta_A+\hat{E}_0)} e^{ -i\pi J_A - i \pi \Delta c } g(\Delta_A+\hat{E}_0).
\end{equation}
We also define a polynomial $c(\hat{E}_0)$ (not to be confused with the central charge) by the formula
\begin{equation}
(\beta \partial_\beta)B(\beta)\bigg|_{\beta=\pi\sqrt{3}} = - |\eta(\omega)|^{-2}\mbox{exp}\{-\pi\sqrt{3} \hat{E}_0 -i\pi\Delta c\}c(\hat{E}_0)
\end{equation}
Using these, we see our differential constraint on the partition function can be expressed as
\begin{equation}
\sum_{A=1}^{\infty}\mbox{exp}\{-\pi\sqrt{3} \Delta_A\} \mbox{exp}\{ -i\pi J_A\} g(\Delta_A+\hat{E}_0)=-c(\hat{E}_0). \label{eq:poly}
\end{equation}

The explicit forms for the defined polynomials are
\begin{gather}
 g(z) = \pi \sqrt{3} z -\frac{1}{2} \nonumber \\
 c(z) = \pi\sqrt{3} z-\frac{1}{2}+\frac{2 \pi \sqrt{3}z}{e^{\pi \sqrt{3}}} - \frac{ 1}{e^{\pi \sqrt{3}}} + \frac{2 \pi \sqrt{3}}{e^{\pi \sqrt{3}}}+\frac{\pi \sqrt{3}z}{e^{
2\pi \sqrt{3}}} - \frac{1}{2e^{2\pi \sqrt{3}}}+\frac{2 \pi\sqrt{3}}{e^{2\pi \sqrt{3}}}.
\end{gather}
In calculating these polynomials, we used the expression
\begin{equation}
\eta'(\omega)=\frac{i\sqrt{3}}{6}\eta(\omega).
\end{equation}
This fact follows from taking a derivative of the modular transformation rule for the Dedekind $\eta$ function
\begin{equation}
\eta\left(\frac{-1}{\tau+1}\right)=\eta(\tau)e^{i\pi/12}\sqrt{\tau + 1}
\end{equation}
and evaluating at $\tau=\omega$.

The polynomials $g(z)$ and $c(z)$ are analogous to the polynomials $f_1(z)$ and $b_1(z)$ from \cite{hell}. Specifically, invariance under the $S$-transformation (using the same assumptions used here) results in the expression
\begin{equation}
\sum_{A=1}^{\infty}\mbox{exp}\{-2\pi \Delta_A\}  f_1(\Delta_A+\hat{E}_0)=-b_1(\hat{E}_0). \label{eq:eq22}
\end{equation}
The polynomials $f_1(z)$ and $b_1(z)$ have the explicit forms
\begin{gather}
 f_1(z) = 2\pi z -\frac{1}{2}\nonumber \\
 b_1(z) = 2\pi z -\frac{1}{2} - \frac{2 (2\pi (z+1) -\frac{1}{2})}{ e^{2\pi} } +\frac{(2\pi (z+2) -\frac{1}{2})}{e^{
4\pi}}.
\end{gather}
In the proof that follows, we will use all four of these polynomial expressions.

\section{Proof of a bound for even spin}
In this section, we will use the polynomial constraints (\ref{eq:poly}) and (\ref{eq:eq22}) to derive a bound on the smallest nonidentity conformal dimension. We wish to proceed using proof by contradiction as in \cite{hell}. The arguments there depend upon having a sum of positive numbers equaling zero. The presence of the complex exponential in equation (\ref{eq:poly}), however, means that terms in our sum could be positive or negative depending on operator spin. To proceed, we will make the assumption that in our theory \emph{all primary operator spins are even}\footnote{Obviously descendant states will include operators with odd spin.}. This is a special property, and so it comes as no surprise that we find tigher bounds than in the case of more general 2D CFTs. There are still many interesting theories that satisfy this assumption, including truncations of pure gravity with scalars, consistent truncations of higher spin gravity theories on AdS$_3$ to massless gauge fields with even spin (and their proposed dual $\mathcal{WD}_N$ minimal model CFTs), and others \cite{j4,j1,j2,j3}.

\subsection{Proof by contradiction}

We now consider a theory with only even-spin primary operators. This restriction eliminates the imaginary part of the exponential in (\ref{eq:poly}), so that the sign of any term in the sum is determined by $g(\Delta_A + \hat{E}_0)$. Having made this assumption, we form the ratio of the lowest order $S$-invariance constraint (\ref{eq:eq22}) and the $ST$-invariance constraint (\ref{eq:poly}):
\begin{equation}
\frac{\sum_{A=1}^{\infty}\mbox{exp}\{-2\pi \Delta_A\}  f_1(\Delta_A+\hat{E}_0)}{\sum_{B=1}^{\infty}\mbox{exp}\{-\pi\sqrt{3} \Delta_B\}  g(\Delta_B+\hat{E}_0)}=\frac{b_1(\hat{E}_0)}{c(\hat{E}_0)}\equiv G_0(\hat{E}_0). \label{eq:ratio}
\end{equation}
Before proceeding, we must address the possibility that eq. (\ref{eq:ratio}) becomes undefined. In Appendix $A$ we demonstrate that $G(\hat{E}_0)$ is defined over the relevant range of central charge and that is strictly positive. 

Subtracting $G(\hat{E}_0)$ over to the RHS, we then combine the terms to get
\begin{equation}
\frac{\sum_{A=1}^{\infty}\left[ e^{-(2-\sqrt{3})\pi \Delta_A}  f_1(\Delta_A+\hat{E}_0) -  g(\Delta_A+\hat{E}_0)G_0(\hat{E}_0)\right] \mbox{exp}\{-\pi\sqrt{3} \Delta_A\}  }    {\sum_{B=1}^{\infty}\mbox{exp}\{-\pi\sqrt{3} \Delta_B\}  g(\Delta_B+\hat{E}_0)}=0. \label{eq:sumrule}
\end{equation}
We now make several definitions in order to simplify our expressions. We define $\alpha \equiv 2 - \sqrt{3}$ and multiply both sides of equation (\ref{eq:sumrule}) by $\exp(-\alpha \pi \hat{E}_0)$ to arrive at
\begin{equation}
\frac{\sum_{A=1}^{\infty}\left[ e^{-(2-\sqrt{3})\pi (\Delta_A+\hat{E}_0)}  f_1(\Delta_A+\hat{E}_0) -  g(\Delta_A+\hat{E}_0)\hat{G}_0(\hat{E}_0)\right] e^{-\pi\sqrt{3} \Delta_A}  }    {\sum_{B=1}^{\infty}\mbox{exp}\{-\pi\sqrt{3} \Delta_B\}  g(\Delta_B+\hat{E}_0)}=0, \label{eq:sumrule2}
\end{equation}
where $\hat{G}_0\equiv G \exp(-\alpha \pi \hat{E}_0)$ (and will be positive by the result of Appendix $A$). We further define the zero of $g$ with respect to $\Delta_A$ as $g^+$. We also define the bracketed expression in the numerator of eq. (\ref{eq:sumrule}) as $P(\Delta_A)$, with the largest root of $P$ labeled as $\Delta^+$. 

We proceed using proof by contradiction; assume $\Delta_1>\mbox{max}(g^+,\Delta^+)$. For positive $\hat{G}$, this implies $P<0$ (as can be checked from the explicit expression) and $g>0$. Because $\Delta_n\geq\Delta_1$ for all $n>1$, we also have that $P(\Delta_n)<0$ and $g(\Delta_n+\hat{E}_0)>0$. Finally, the reality of $\Delta_n$ implies exp$\{-\pi\sqrt{3}\Delta_i\}>0$ for $i\geq1$. Thus the denominator is always positive and every term in the numerator is negative for $\Delta_1>\mbox{ max}(g^+,\Delta^+)$. It is impossible to add together negative numbers to equal zero: we therefore have a contradiction. We have thus derived our first bound:
\begin{equation}
\Delta_1\leq \mbox{ max}(g^+,\Delta^+).
\end{equation}

Using the explicit form of $g(z)$, we can find an exact expression for $g^+$:
\begin{equation}
g^+=\frac{c_{\rm tot}}{24}+\frac{\sqrt{3}}{6\pi}-\frac{1}{12}\approx \frac{c_{\rm tot}}{24}+0.00855482...
\end{equation}
In order to simplify our bound, we now turn our attention to the root $\Delta^+$.

\subsection{Analytic and numerical bounds on $\Delta_1$}

In this section, we find analytic and numerical upper bounds on $\Delta^+$. We do so without reference to asymptotically large central charge, which results in truly universal bounds in this class of theories. We begin by considering the explicit expression for $P=0$,
\begin{equation}
\left[2\pi(\Delta^++ \hat{E}_0) -\frac{1}{2}\right]e^{-\alpha\pi(\Delta^++\hat{E})}-\left[ \pi\sqrt{3}(\Delta^++ \hat{E}_0)  -\frac{1}{2}\right]\hat{G}_0(\hat{E}_0)=0. \label{eq:ahs}
\end{equation}
To simplify analysis, we define $z^+\equiv \pi(\Delta^+ + \hat{E}_0 )$. Then eq. (\ref{eq:ahs}) becomes
\begin{equation}
\left( \sqrt{3}z^+ - \frac{1}{2}\right)\hat{G}_0=\left(2z^+-\frac{1}{2}\right)e^{-\alpha z^+}.   \label{eq:zee}
\end{equation}
We will use this expression to bound $\Delta^+$.

Due to sign considerations, $z^+$ can only exist on the intervals $z^+<\frac{1}{4}$ and $z^+>\frac{\sqrt{3}}{6}$. 
We consider first the latter interval. The positivity of $z^+$ on this interval means
\begin{eqnarray}
\left( \sqrt{3}z^+ - \frac{1}{2}\right)\hat{G}_0=\left(2z^+-\frac{1}{2}\right)e^{-bz^+}<\left(2z^+-\frac{1}{2}\right) \nonumber \\
\Rightarrow z^+ < \frac{\hat{G}-1}{2\sqrt{3}\hat{G}-4} \label{eq:gbound}
\end{eqnarray}
In performing this simplification, I have assumed that $\hat{G}>2\sqrt{3}/3$. This is equivalent to the condition that  $c_{\rm tot} > 2.33544...$, which is a stronger assumption than $c_{\rm tot}> 2$. We will address this further restriction momentarily. As $\hat{G}$ approaches $\frac{2\sqrt{3}}{3}$, the RHS approaches $+\infty$ and we can prove no bound. As we increase $\hat{G}$, the RHS monotonically decreases and asymptotes to $\frac{\sqrt{3}}{6}$. 

Thus the possibilities are that $z^+<\frac14$---in which case equation (\ref{eq:gbound}) trivially holds---or $z^+ > \frac{\sqrt{3}}{6}$ in which case equation (\ref{eq:gbound}) also holds. In either case, we have proven
\begin{equation}
z^+=\pi(\Delta_+ + \hat{E}_0) < \frac{\hat{G}-1}{2\sqrt{3}\hat{G}-4} \Rightarrow \Delta_+ < \frac{c_{\rm tot}}{24} + \frac{\hat{G}-1}{2\sqrt{3}\hat{G}-4} - \frac{1}{12}. \label{eq:steps}
\end{equation}
or upon simplifying
\begin{equation}
\Delta_+ < \frac{c_{\rm tot}}{24} +  \frac{\sqrt{3}}{6\pi} - \frac{1}{12}  + \frac{\alpha/\pi}{6\hat{G}-4\sqrt{3}} .
\end{equation}
In this form, it is clear that $\mbox{max}(\Delta^+,g^+)=\Delta^+$, and we therefore have the bound
\begin{equation}
\Delta_1 \leq \Delta^+ = \frac{c_{\rm tot}}{24}  +  \frac{\sqrt{3}}{6\pi} - \frac{1}{12}  + \frac{\alpha/\pi}{6\hat{G}-4\sqrt{3}}\label{eq:bound}.
\end{equation}
This bound holds for all two-dimensional conformal field theories with only even spin primary operators, subject to the constraints that there are no chiral algebras beyond the Virasoro algebra and that $c> 1$, $\tilde{c}>1$, and $c_{\rm tot}>2.33544..$.

It is apparent that tighter bounds can be calculated by restricting the allowed values of the total central charge. For example, if we restrict ourselves to the case where $c_{\rm tot} \geq 2.5$, an explicit calculation gives that
\begin{equation}
c_{\rm tot} \geq 2.5 \Rightarrow \Delta_1 \leq \Delta^+ = \frac{c_{\rm tot}}{24}+ 2.1510...  \nonumber
\end{equation}
Additional calculations give
\begin{eqnarray}
c_{\rm tot} \geq 3 \Rightarrow \Delta_1 \leq \Delta^+ = \frac{c_{\rm tot}}{24}+ 0.5338..., \nonumber \\
c_{\rm tot} \geq 4 \Rightarrow \Delta_1 \leq \Delta^+ = \frac{c_{\rm tot}}{24}+ 0.2142..., \nonumber \\
c_{\rm tot} \geq 48 \Rightarrow \Delta_1 \leq \Delta^+ = \frac{c_{\rm tot}}{24}+ 0.0116... \nonumber \\
\end{eqnarray}        
In the limit of asymptotically large total central charge, the numerical constant converges to $\frac{\sqrt{3}}{6\pi} - \frac{1}{12}  \approx 0.008554...$

Although the bound (\ref{eq:bound}) is analytic, it is not without its deficiencies. First, there is no reason to expect that it is optimal in the sense that it is saturated by some physical conformal field theory. Indeed, the argument (\ref{eq:steps}) was formulated due to its convenience; a more careful analysis should give a significantly tighter bound. Second, our particular derivation further restricted the allowed range of total central charge. Though this is not cataclysmic, a more general bound valid for the full range of central charge $c,\tilde{c} > 1$ is preferred. To find such an improved bound, therefore, we proceed numerically.

The largest root $\Delta^+$ (or analogously, $z^+$) of the polynomial $P$ satisfies equation (\ref{eq:zee}). We seek the least upper bound on $z^+$ for $c_{\rm tot}>2$ The function $z^+$ attains a global maximum (for $c_{\rm tot} \approx 2$), so that
\begin{equation}
z^+ < 0.5530...
\end{equation}
Substituting the definition of $z$ thus gives the numerical bound
\begin{equation}
\Delta_1 < \frac{c_{\rm tot}}{24} + 0.09270...
\end{equation}
This is a notable improvement over the bounds determined analytically. As in that case, restricting the central charge to larger values gives a tighter bound. For example,
\begin{equation*}
c_{\rm tot} \geq 48 \Rightarrow \Delta_1 < \frac{c_{\rm tot}}{24} + 0.00903...
\end{equation*}
Again, in the limit of asymptotically large total central charge the numerical constant converges to $\frac{\sqrt{3}}{6\pi} - \frac{1}{12}  \approx 0.008554...$

\section{Gravitational interpretation of bounds}

In this section, we briefly explore the gravitational interpretation of our CFT results using the AdS/CFT correspondence. Our restriction to CFTs with only even spin operators means that the corresponding gravitational theory must also have only even spin primary operators. We therefore restrict our discussions to the relevant gravitational duals---even spin truncations of gravitational or higher-spin gravitational theories. In the case of AdS$_3$/CFT$_2$, the matching between the central charge of the CFT and the cosmological constant identified in \cite{26p} is
\begin{equation}
c+\tilde{c}=\frac{3}{G_N \sqrt{|\Lambda|}},
\end{equation}
where $\Lambda=-L^{-2}$ and $L$ is the AdS radius. From this expression, it's clear that the flat space limit corresponds to taking the limit $c_{\rm tot}\rightarrow \infty$.
We also match primary operators\footnote{A primary state corresponds to a state at rest, and descendant states correspond to the original massive state in the bulk with boundary metric excitations \cite{6p}.} with some conformal dimension living in the boundary CFT with massive objects in the bulk with some center-of-mass energy according to the identification 
\begin{equation}
E^{com}=\frac{\Delta}{L}.
\end{equation}
It is clear that in the flat space limit, only terms proportional to or larger than the total central charge will contribute to the mass of the bulk state.

With these identifications, equation (\ref{eq:bound}) says that every suitable theory of quantum gravity having only even spin fields must have a massive state in the bulk of rest energy $M_1$ such that
\begin{equation}
M_1\leq M_+\equiv\frac{1}{L}\Delta^+|_{c_{total}=\frac{3L}{G_N}}.
\end{equation}
Using our analytic expression for $\Delta^+$, this inequality becomes
\begin{equation}
M_1\leq \frac{1}{8G_N}+\frac{d_0}{L}
\end{equation}
where $d_0\equiv +  \frac{\sqrt{3}}{6\pi} - \frac{1}{12}  + \frac{\alpha/\pi}{6\hat{G}-4\sqrt{3}}$. For our allowed values of the central charge ($c_{\rm tot}>2.3354...$), the RHS of this bound is finite. Similar results apply for the case where we use our numerical expressions for $\Delta^+$. This restriction to $c,\tilde{c} > 1$ (or $c_{\rm tot}>2.3354... $) is not overly restrictive, as the range $0 < c_{\rm tot} \lesssim 2$ represents AdS$_3$ spaces with Planck-scale curvatures. Theories of gravitation with such extreme curvatures are exotic, at best. In the flat-space limit $\Lambda\rightarrow0$, this bound becomes
\begin{equation}
M_1\leq \frac{1}{8G_N}.
\end{equation}
This mass is precisely the rest energy of the lightest BTZ black hole. Because any theory of 3D gravity admitting an AdS vacuum will also admit BTZ black holes, we interpret this bound as saying that there should always be a massive state in such even-spin theories at around the energy scale corresponding to the lightest spinless BTZ black hole. \cite{27p,28p}

\appendix
\section{Properties of $G_0(\hat{E}_0)$}

In this section, we demonstrate that $\hat{G}_0(\hat{E}_0)>0$ for the interval $\hat{E}_0 \in (-\infty,0)$ (corresponding to values of the total central charge  $c_{\rm tot}\in (2,\infty)$ ). The exponential factor is obviously positive for real arguments, so we need only prove ${G}_0(\hat{E}_0)>0$. This function has one simple pole at
\begin{equation}
 \hat{E}_0 =  \frac{1}{6}  \frac{\sqrt{3}e^{\pi\sqrt{3}}-12\pi +\sqrt{3}}{\pi(e^{\pi\sqrt{3}}+1)} \approx 0.083257...
\end{equation}
This point is outside of our interval for the central charge, and so the function ${G}_0$ will be continuous for $c_{\rm tot} > 2$. 

Furthermore, ${G}_0$ has only one zero, at
\begin{equation}
\hat{E}_0 = -\frac{1}{4}\frac{1+8\pi e^{-2\pi}-2 e^{-2\pi}-8\pi e^{-4\pi}+ e^{-4\pi}}{\pi(-1+2 e^{-2\pi} - e^{-4\pi})}\approx 0.083319...
\end{equation}
This point is also outside of our interval of interest. Thus for relevant values of the central charge, the function $G_0(\hat{E}_0)$ will be either everywhere positive or everywhere negative. 

To determine whether $G_0(\hat{E}_0)$ is strictly negative or positive, we need only calculate its value for a specific value of $\hat{E}_0$. Evaluating at $c=42$ gives $G_0\approx 1.14050..$, and therefore $\hat{G}_0(\hat{E}_0)>0$.

\section*{Acknowledgments}

I would like to thank Al Shapere for his assistance in preparing this work and Diptarka Das for helpful discussions. I would like to thank the JHEP referee(s) for their helpful feedback. I also thank the University of Kentucky and National Taiwan University for their financial support. This work was supported in part by grants NSF \#0970069 and a University of Kentucky fellowship.

\end{document}